\documentclass[11pt]{article}
\usepackage[utf8]{inputenc}%

\usepackage[tracking=true, letterspace=100, expansion=false]{microtype}
\usepackage[T1]{fontenc}
\usepackage{amsmath}%
\usepackage{amsfonts}%
\usepackage{amsthm}%
\usepackage{booktabs}
\usepackage[numbers]{natbib}
\usepackage{tabularx}
\usepackage{caption}
\usepackage[labelformat=simple]{subcaption}
\usepackage{multirow}
\usepackage{graphicx}

\theoremstyle{definition}

\theoremstyle{remark}

\usepackage[margin=1.2in]{geometry}
\usepackage[onehalfspacing]{setspace}
\usepackage{mathtools}

\usepackage{bbm} 

\usepackage[title]{appendix}
\usepackage{etoolbox} 
\AtBeginEnvironment{appendices}{\crefalias{section}{appendix}}

\usepackage{changepage}

\usepackage{xcolor}%
\definecolor{webbrown}{rgb}{.6,0,0}%
\usepackage{hyperref} %
\hypersetup{%
  breaklinks = true,%
  colorlinks = true,%
  anchorcolor = webbrown,%
  citecolor = webbrown,%
  filecolor = webbrown,%
  linkcolor = webbrown,%
  menucolor = webbrown,%
  urlcolor= webbrown,%
  citebordercolor= 1 0 0,%
  menubordercolor=1 0 0,%
  urlbordercolor=1 0 0,%
  runbordercolor=1 0 0,%
  pdftitle={Excess death rates for Republicans and Democrats during the COVID-19 pandemic},
  pdfauthor={Wallace, Goldsmith-Pinkham, and Schwartz}
}

\usepackage{cleveref}
\crefname{appsec}{appendix}{appendices}
\crefname{appsubsec}{appendix}{appendices}
\crefname{assumption}{assumption}{assumptions}

\usepackage[nolist]{acronym}
\begin{acronym}
  \acro{CI}{confidence interval}%
  \acro{OLS}{ordinary least squares}%
  \acro{CLT}{central limit theorem}%
  \acro{IV}{instrumental variables}%
  \acro{ATE}{average treatment effect}%
  \acro{RCT}{randomized control trial}%
  \acro{VAM}{value-added model}%
  \acro{LAN}{locally asymptotically normal}%
  \acro{DiD}{difference-in-differences}%
  \acro{OVB}{omitted variables bias}
  \acro{FWL}{Frisch-Waugh-Lovell}
\end{acronym}

\title{Excess death rates for Republicans and Democrats during the COVID-19 pandemic\thanks{We gratefully acknowledge support from the Tobin Center for Economic Policy at Yale University and the Yale School of Public Health COVID-19 Rapid Response Research Fund. }}

\author{Jacob Wallace\thanks{Yale School of Public Health  Email: \href{mailto:jacob.wallace@yale.edu}{jacob.wallace@yale.edu}} \and Paul Goldsmith-Pinkham\thanks{Yale School of Management and NBER Email: \href{mailto:paul.goldsmith-pinkham@yale.edu}{paul.goldsmith-pinkham@yale.edu}} \and Jason L. Schwartz\thanks{Yale School of Public Health Email: \href{mailto:jason.l.schwartz@yale.edu}{jason.l.schwartz@yale.edu}}}
\date{\today}

\begin{document}
\begin{titlepage}
  \maketitle
  \thispagestyle{empty}

\begin{adjustwidth*}{0.1cm}{0.1cm}
\begin{abstract}
 Political affiliation has emerged as a potential risk factor for COVID-19, amid evidence that Republican-leaning counties have had higher COVID-19 death rates than Democrat-leaning counties and evidence of a link between political party affiliation and vaccination views. This study constructs an individual-level dataset with political affiliation and excess death rates during the COVID-19 pandemic via a linkage of 2017 voter registration in Ohio and Florida to mortality data from 2018 to 2021. We estimate substantially higher excess death rates for registered Republicans when compared to registered Democrats, with almost all of the difference concentrated in the period after vaccines were widely available in our study states. Overall, the excess death rate for Republicans was 5.4 percentage points (pp), or 76\%, higher than the excess death rate for Democrats. Post-vaccines, the excess death rate gap between Republicans and Democrats widened from 1.6 pp (22\% of the Democrat excess death rate)  to 10.4 pp (153\% of the Democrat excess death rate). The gap in excess death rates between Republicans and Democrats is concentrated in counties with low vaccination rates and only materializes after vaccines became widely available.
\end{abstract}
\end{adjustwidth*}
\end{titlepage}

Coronavirus disease 2019 (COVID-19) has caused over one million deaths in the United States \citep{cdc}, leading to considerable interest in identifying risk factors for COVID-19 mortality. Political affiliation has emerged as one potentially significant risk factor, amid evidence that Republican-leaning counties have had higher COVID-19 death rates than Democrat-leaning counties \citep{leonhardt,sehgal2022association}. But it is unknown whether this county-level association — which may be subject to the ecological fallacy \citep{piantadosi1988ecological} — persists at the individual level. Prior research has also established differences in vaccination attitudes and social distancing based on political party affiliation \citep{allcott2020polarization, grossman2020political, callaghan2021correlates, cowan2021covid, pink2021elite}, but it has been more difficult thus far to establish corresponding links to health outcomes due to data limitations. This study overcomes that challenge by linking voter registration data in Ohio and Florida to mortality data to assess the individual-level association between political party affiliation and excess mortality during the COVID-19 pandemic. We estimate higher excess death rates for registered Republicans when compared to registered Democrats after vaccines were widely available — and not before — and these differences were concentrated in counties with lower vaccination rates.

\section*{Results}

To calculate excess deaths, we use 577,659 deaths of individuals linked to their 2017 voting records in Ohio and Florida who died at age 25 or older between January 2018 and December 2021. Our approach estimates ``excess death rates'' as the percent increase in deaths above expected deaths that are due to seasonality, geographic location, party affiliation, and age. These expected deaths are calculated non-parametrically using 2019 data by aggregating deaths into counts $N_{mcpa,2019}$ at the month-by-county-by-party-by-age-bin level. The age bins used were 25-64, 65-74, 75-84, and 85-and-older. We then estimate relative excess deaths $E_{mcpa,t}$ in other periods $t$ for a given county-by-party-by-age cell by considering the percent change in deaths: 
\begin{equation}
    E_{mcpa,t} = (N_{mcpa,t} - N_{mcpa,2019})\big/N_{mcpa,2019}.
\end{equation}
We also report aggregated excess death measures (e.g., the percent increase in excess deaths in 2020 and 2021) by taking the average of $E_{mcpa,t}$, weighted by $N_{mcpa,2019}$. Thiscontrols for differences in \emph{pre}-COVID-19 death counts across calendar month, county of residence at time of voter registration, political party registration (Democrat or Republican), and age bins.

\begin{figure}
\centering
\includegraphics[width=\linewidth]{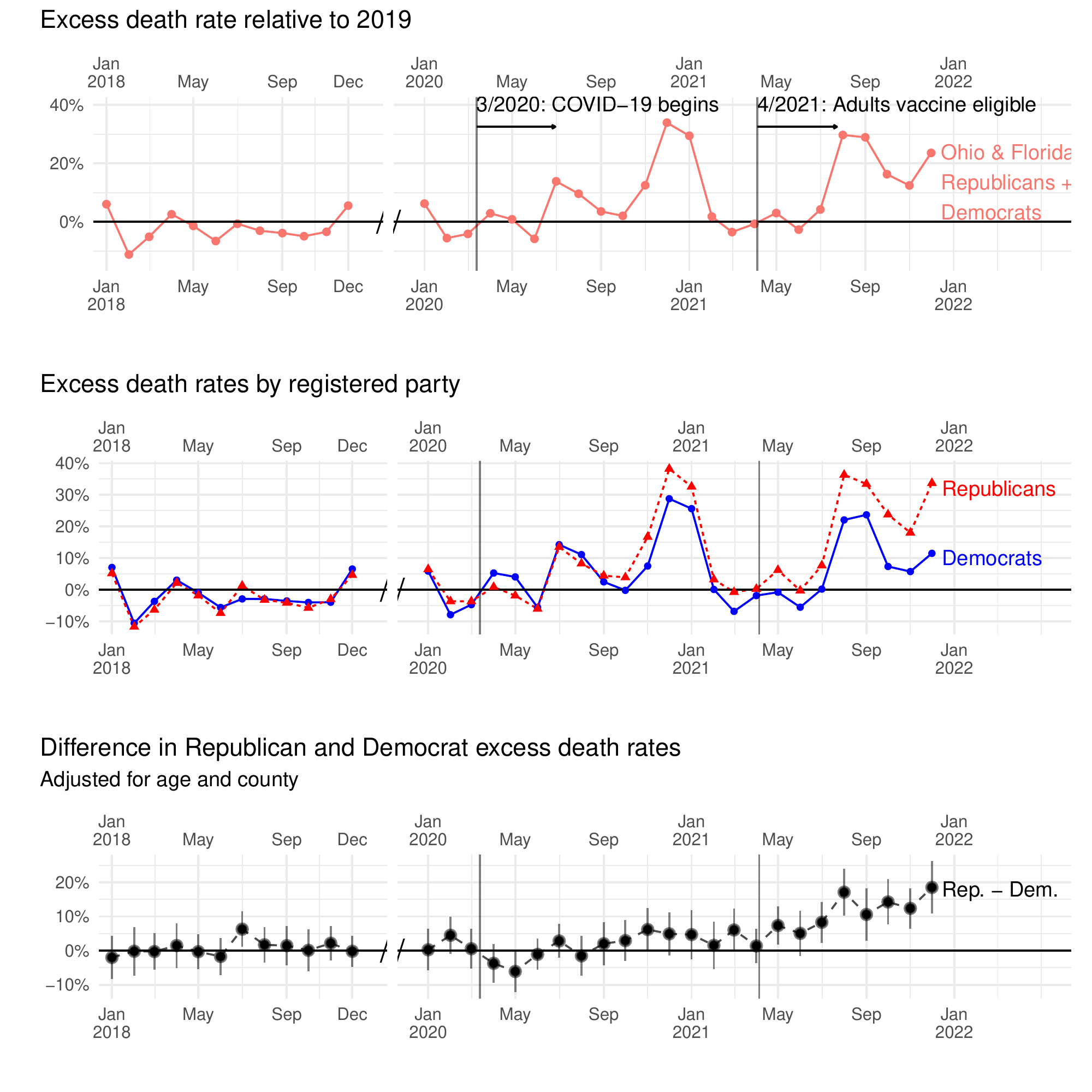}
\caption{The figure plots monthly excess deaths for Florida and Ohio based on mortality records linked to voter registration files. The first panel plots aggregate excess death rates. The second panel plots excess death rates separately for registered Republicans and Democrats. The third panel plots the percentage point difference in excess deaths between Republicans and Democrats after regression adjusting for year-month-by-age-bin-by-county differences. The error bars reflect the pointwise 95\% confidence interval.}
\label{fig:main_deaths}
\end{figure}

In Figure \ref{fig:main_deaths}, we plot the values of $E_{mcpa,t}$ by months for 2018, 2020, and 2021 in different ways. In the first panel we report the percent excess deaths in aggregate, pooling registered Democrats and Republicans. The aggregate excess death rate from March 2020 to December 2021 was 10.3 percentage points (pp) (95\% CI:6.7,14.0) (\ref{fig:main_deaths}). Calendar year 2018 and the beginning of 2020 serve as a placebo test for our method, as excess death rates should be approximately zero prior to the COVID-19 pandemic. As expected, Figure \ref{fig:main_deaths} depicts only small fluctuations in excess death rates around zero prior to the pandemic. Finally, the qualitative patterns of excess deaths we observe in Florida and Ohio during this time match the trends reported elsewhere, but the magnitudes are slightly smaller \citep{cdc2}.

In the second panel of Figure \ref{fig:main_deaths}, we plot excess death rates for Republicans and Democrats separately. In 2018 and the early parts of 2020, excess death rates for Republicans and Democrats are similar, and centered around zero. Both groups experienced a similar large spike in excess deaths in the winter of 2020-2021. However, in the summer of 2021 — after vaccines were widely available — the Republican excess death rate rose to nearly double that of Democrats, and this gap widened further in the winter of 2021.  

In the third panel of Figure \ref{fig:main_deaths}, we estimate the difference between Republican and Democrat excess death rates by regressing $E_{mcpa,t}$ on Republican-year-month indicators, controlling for year-month-by-county-by-age-bin fixed effects. These controls adjust for differences in excess death rates \emph{during} COVID-19, in addition to the adjustments for death rates by age, party and location from the construct of excess death. We report the point estimates and 95\% confidence intervals for each of these estimates (standard errors are clustered by county). A joint test of the null hypothesis that the coefficients in 2018 and January and February of 2020 are equal to zero fails to reject  ($p$ = 0.62). Between March 2020 and December 2021, the part of our study period that overlaps the COVID-19 pandemic, average excess death rates were 5.4 pp (76\%) higher among Republicans than Democrats (95\% CI:2.9,8.0; $p\leq0.001$). However, this gap widened from 1.6 pp (95\% CI:-1.2,4.4; $p=0.25$) between March 2020 and March 2021, to a 10.6 pp (95\% CI:7.3,13.8; $p\leq0.001$), or 153\%, difference after April 5, 2021, when all adults were eligible for COVID-19 vaccines in Florida and Ohio.

This sharp contrast in the excess death rate gap before and after vaccines were available suggests that vaccine take-up likely played an important role. Data on vaccine take-up by party is limited and unavailable in our dataset, but there is evidence of differences in vaccination attitudes and reported uptake based on political party affiliation \citep{pink2021elite,kff,civiqs}. Using county-level vaccination rates, we find evidence that vaccination contributes to explaining differences in excess deaths by political party affiliation, even after controlling for location and age differences.

\begin{figure}[t]
\centering
\includegraphics[width=\linewidth]{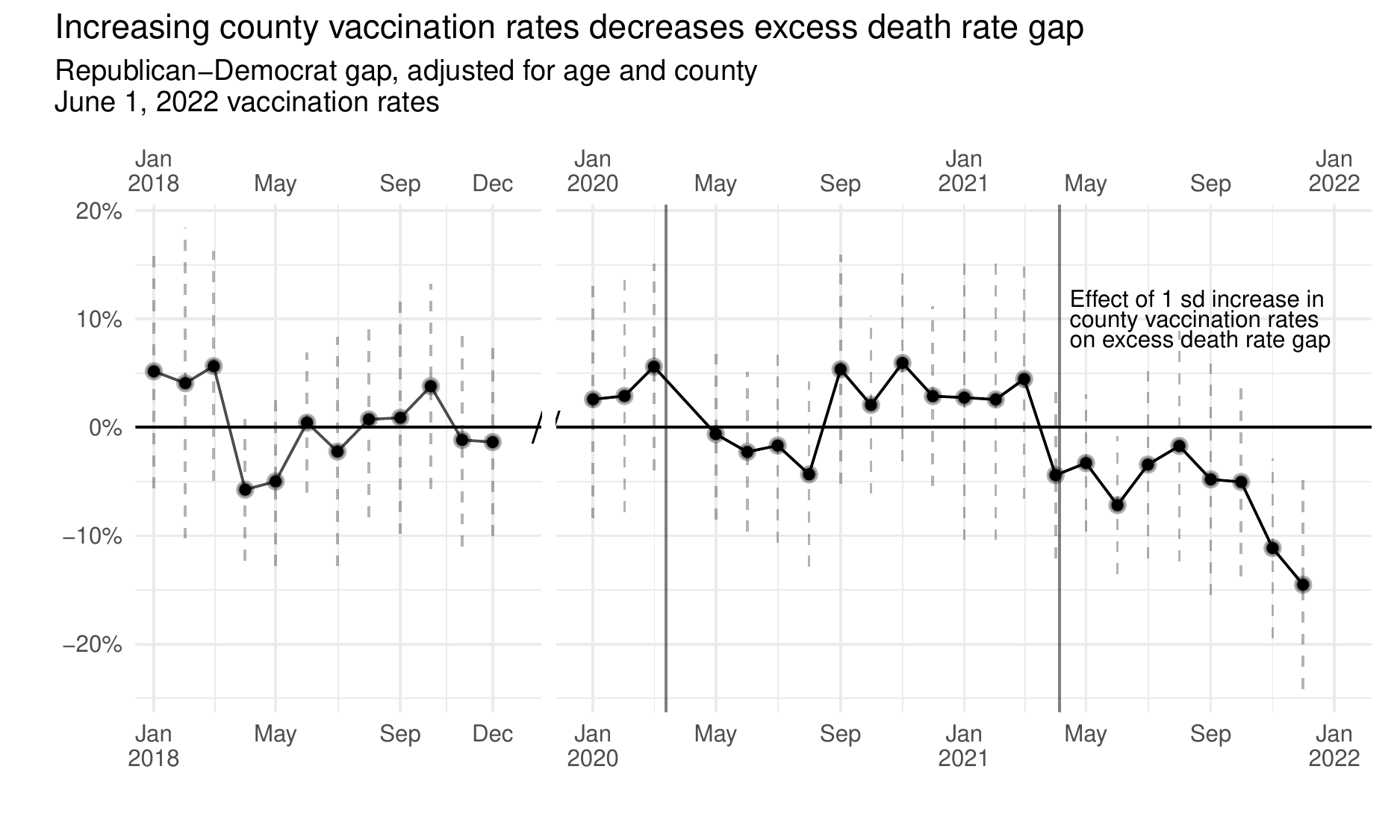}
\caption{The figure plots the association between a one standard deviation increase in the county-level vaccination rate (roughly 10 percent) on the Republican-Democrat monthly excess death rate gap in Florida and Ohio. The coefficients regression adjust for year-month-by-age-bin-by-county differences. The error bars report the 95\% confidence interval. }
\label{fig:vaccine_event}
\end{figure}

We reexamine the Republican-Democrat excess death rate gap from Figure \ref{fig:main_deaths}, but now interact Republican-year-month indicators with a measure of county-level vaccination rates (the number of individuals with at least one dose scaled by population) as of June 1, 2021. This measure is scaled to have a standard deviation (SD) of one, so the coefficient is the association between a one SD increase in the county-level vaccination rate (roughly 10 percent) and the magnitude of the Republican-Democrat difference in excess death rates. In Figure \ref{fig:vaccine_event}, we report these monthly coefficients over time, along with 95\% confidence intervals. We do not observe a statistically significant association between the county-level vaccination rate and the Republican-Democrat excess death gap until after the vaccine is widely available, with a sharp (and increasingly) negative effect toward the end of sample. This suggests that during the period after vaccines were widely available, areas with larger take-up of vaccines saw much smaller excess death rate gaps between Republicans and Democrats (even after adjusting for county and age differences).

\begin{figure}[t]
\centering
\includegraphics[width=0.8\linewidth]{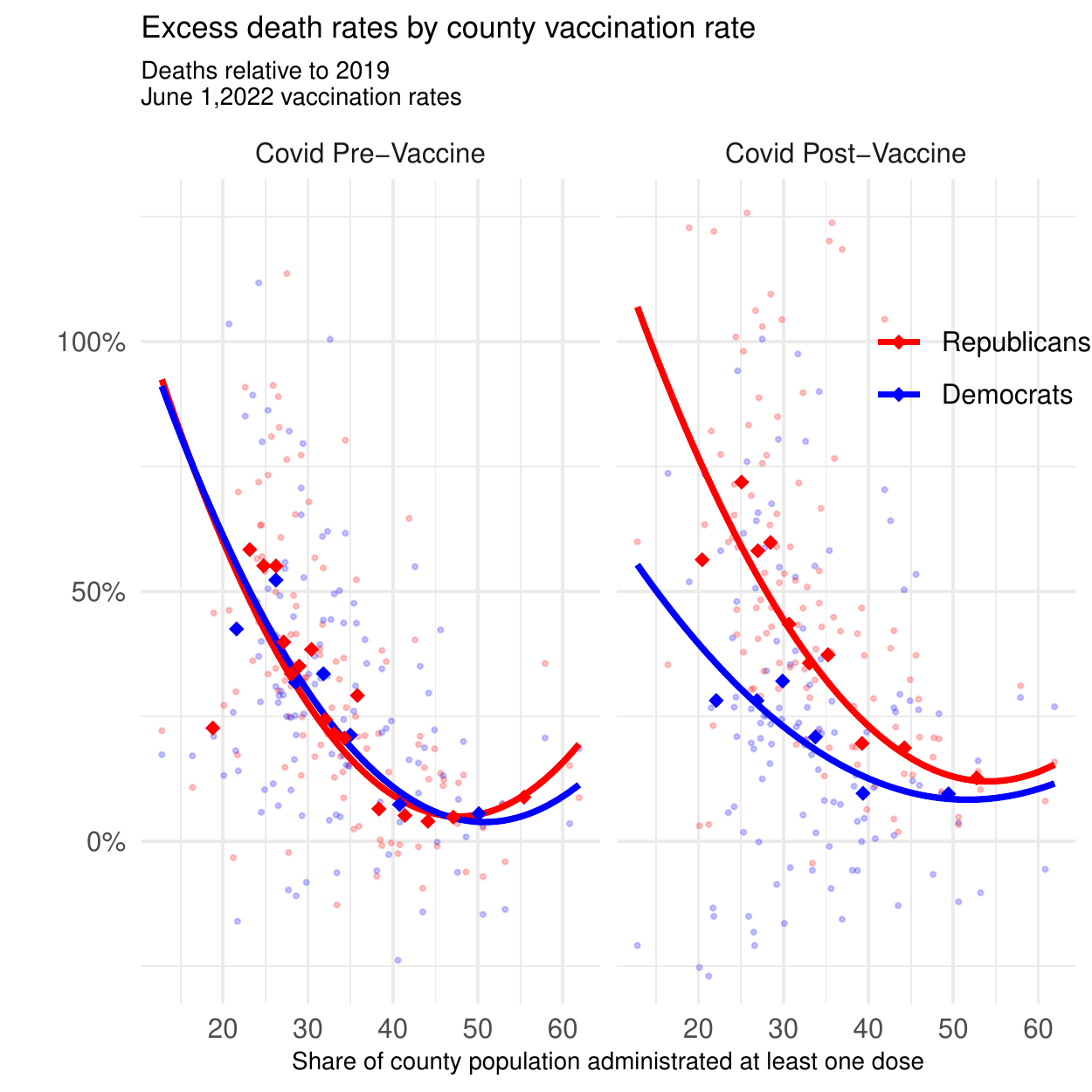}
\caption{The figure plots average excess death rates for Republicans and Democrats in each county for Covid Pre-Vaccine (April 2020 to March 2021) and Covid Post-Vaccine (April 2021-December 2021). The circles are each county, and diamonds are binned means \citep{cattaneo2019binscatter} for each party. The curves are quadratic fits using least squares.}
\label{fig:vaccine_county}
\end{figure}

In Figure \ref{fig:vaccine_county}, we explore the association between county-level vaccination rates and excess death rates for Republicans and Democrats before and after vaccines were widely available. We pool our excess death data into two time periods: Covid ``Pre-Vaccine'' (April 2020 to March 2021) and Covid ``Post-Vaccine'' (April 2021-December 2021), referring to the period when vaccines were available to all adults in Florida and Ohio. For each time period, we calculate the average excess death rates for Republicans and Democrats in each county, and then calculate binned means of excess death rate \citep{cattaneo2019binscatter} for each party. Finally, we fit quadratic curves for each group using least squares. 

Two noteworthy facts emerge in Figure \ref{fig:vaccine_county}. First, in the Covid Pre-Vaccine period, the association between excess death rates and county-level vaccination rates are nearly identical for Democrats and Republicans. Statistically, we fail to reject differences in both slopes and intercept between the two groups in this period ($p$ = 0.73 for slopes and $p$ = 0.84 for slopes and intercepts). Second, in the Covid Post-Vaccine period, there is a clear visual difference between Democrats and Republicans ($p$ =0.02 for slopes and $p$ = 0.00 for slopes and intercepts), with higher excess death rates for Republicans in counties with lower vaccination rates. By comparison, the difference in excess deaths between Republicans and Democrats is nearly zero in  counties with the highest vaccination rates.

\section*{Discussion }

Political party affiliation was associated with excess death rates at the individual level during the initial years of the COVID-19 pandemic. Registered Republicans in Florida and Ohio had higher excess death rates than registered Democrats, driven by a large mortality gap in the period after all adults were eligible for vaccines. These results adjust for county-by-age differences in excess deaths during the pandemic, suggesting that there were within-age-by-county differences in excess death associated with political party affiliation.

Our study has several limitations. First, our mortality data, while detailed and recent, only includes approximately 80\% of deaths in the US. However, excess death patterns in our data are similar to those in other reliable sources. Second, because we did not have information on an individual's vaccination status, analyses of the association between vaccination rates and excess deaths relied on county-level vaccination rates. Third, our study is based on data from the only states where we could obtain voter registration information (Florida and Ohio); hence, our results may not generalize to other states.

Overall, our results suggest that political party affiliation only became a substantial risk factor in Ohio and Florida after vaccines were widely available. Lack of individual-level vaccination status limits our ability to draw broad conclusions, but the results suggest that the well-documented differences in vaccination attitudes and reported uptake between Republicans and Democrats \citep{kff,civiqs,cowan2021covid, pink2021elite} have already had serious consequences for the severity and trajectory of the pandemic in the United States. If these differences in vaccination by political party affiliation persist, then the higher excess death rate among Republicans is likely to continue through the subsequent stages of the COVID-19 pandemic.

\section*{Materials and Methods}

\paragraph{Data} Detailed mortality data for 2018-2021 was obtained from Datavant, an organization that uses privacy-preserving record linkage to connect the Social Security Administration Death Master File with information from newspapers, funeral homes, and other sources to construct an individual-level database with over 80\% of annual US deaths. This data was linked at the individual level to 2017 Florida and Ohio voter registration files \citep{dataverse}. For each record, the linked data include month of death, age of deceased, county of residence, and 2017 political party registration. Political party affiliation in Florida is actively registered by voters. Political party affiliation in Ohio is defined by whether they voted in a party’s primary election within the preceding two calendar years. We excluded individuals that were registered independents or affiliated with third parties. We obtained county-level vaccination rates as of June 1, 2021 from the Centers for Disease Control and Prevention \citep{cdc_vaccine}.

 A web appendix, aggregated data, and analysis scripts will be deposited in the Open Science Framework (\url{https://osf.io/m42d7/}). Researchers interested in using individual-level data will need to acquire the mortality data from Datavant. The study was deemed exempt by the Yale University IRB.

\paragraph{Statistical measurement} To construct an estimate of excess death rates, we aggregated death counts at the month-by-county-by-party-by-age level. The age bins used were 25-64, 65-74, 75-84, and 85+. We calculated percent increases in deaths by calendar month in 2018, 2020 and 2021, relative to 2019, for each county-by-party-by-age cell. When plotting aggregated excess death rates by month, we take averages weighted by baseline death counts. We then estimated differences in excess death rates by period and political party affiliation, plotted them, and assessed their association with county-level vaccination rates. Statistical models adjusted for differences in excess death rates by calendar quarter, county, and age.

Measurement error does not appear to affect our conclusions of the excess death rate gap. First, in Figure \ref{fig:main_deaths} the gap in rates is statistically indistinguishable from zero in 2018 and the beginning of 2020, suggesting no correlation between any measurement error and party affiliation. Second, to the extent we undermeasure deaths during the COVID-19 period and it is uncorrelated with party affiliation, then we still correctly estimate the relative \emph{percentage} gap of 76 percent.

\setstretch{1.15}

\bibliographystyle{abbrv}
\bibliography{bibliography.bib}

\clearpage

\begin{appendix}

  \begin{center}
    \Large
Supplementary Information for\\

``Excess death rates for Republicans and Democrats during the COVID-19 pandemic''

\end{center}
\section{Supplemental description of methods and data}
\subsection{Description of study data and linkages}

This study made use of data from four different sources. Below, we describe each of these data sources in more detail and then discuss how these data were assembled into our analytic file. Portions of the dataset will be publicly available on an OSF repository upon publication: https://osf.io/m42d7/
These include replication code using aggregated data, aggregated data for our two main datasets used in the paper, the Ohio Voter File, and this supplement.

\paragraph{Florida voter file:} The publicly available Florida voter file for February 2017 was accessed via the Harvard Dataverse. For additional details on the file and a link to request access to the data proceed to the following: \url{https://dataverse.harvard.edu/dataset.xhtml?persistentId=doi:10.7910/DVN/UBIG3F}. The file contains full name, date of birth, county of registration, gender, and party affiliation.

\paragraph{Ohio voter file:} The publicly available Ohio voter file for 2017 was accessed via the Ohio Secretary of State website at: \url{https://www6.ohiosos.gov/ords/f?p=VOTERFTP:HOME}. We accessed the link and downloaded the data on March 4, 2017. The file contains full name, date of birth, county of registration, and party affiliation.

\paragraph{Datavant}:
Detailed mortality data for 2018 to 2021 was obtained from Datavant, an organization that uses privacy-preserving record linkage to connect the Social Security Administration Death Master File with information from newspapers, funeral homes, and other sources to construct an individual-level database with over 80\% of annual US deaths. For each record, the identifiable data indicate the month of death and age of the deceased individual in months, as well as individual-level identifiers that could be used to link to the voter files but that we, as researchers, did not have access to in the linked data. The Datavant mortality data was then linked, at the individual level, to the Florida and Ohio voter files on first name, last name, and date of birth using a proprietary algorithm. A match indicated that a registered voter in Florida and Ohio had been identified in the Datavant mortality data and provided detailed information on date of death and the age of the deceased individual in months, top-coded at 89 years old. A deidentified version of the linked data was then provided to our research team. That dataset contained date of death, year-month of birth, county of residence based on the voter registration file in 2017, party affiliation, and either gender (if from Florida) or a probabilistic guess at gender (if from Ohio) based on first name. 

\paragraph{Center for Disease Control and Prevention (CDC):}
We obtained additional information from the CDC. From the CDC, we also accessed data on county-level vaccination rates as of June 23, 2022. That data can be accessed here: \url{https://data.cdc.gov/Vaccinations/COVID-19-Vaccinations-in-the-United-States-County/8xkx-amqh}. From that data, we obtained information on what share of a county’s population had received at least one dose of a Covid-19 vaccine as of June 6, 2021. We selected this date—two months after vaccines became available to all adults in our study states—because it represented the approximate time when all adults had had the opportunity (if they so desired) to complete the two-dose vaccine series as well as the additional 14 days following the second dose needed to be considered “fully vaccinated” according to the CDC.

\subsection{Methodology for calculating excess death rates}

To construct an estimate of excess death rates, we first aggregated death counts at the month-by-county-by-party-by-age level. The age bins that we used were: 25-64, 65-74, 75-84, 85+. We kept registered Republicans and Democrats and removed all unaffiliated or independent voters, as well as voters registered to any other political parties.

For each month in 2018, 2020, and 2021, we then calculated the percent increase in deaths relative to 2019. While our aggregated data is at the month-by-county-by-party-by-age level, we map months to quarters (e.g., January-March to quarter 1) to coarsen our cells and reduce noise in estimating the expected monthly death count from which excess deaths in 2018, 2020, and 2021 are constructed. Effectively, this means that all the months in a quarter have the same predicted death count conditional on a particular county-by-party-by-age cell. For example, we would measure the excess death rate for 25-64 year old Republicans registered in Franklin county (in Ohio) in each month of 2020 by dividing the count of deaths for individuals in that group by the average monthly death count in the first quarter of 2019 for that age cell (25-64 year old Republicans in Franklin county in Ohio). Intuitively, this measure of excess deaths adjusts for baseline (i.e., pre-COVID-19) differences in mortality by county, registered political party, calendar quarter, and age bins. When we plot the aggregated data for a particular month (Figure), we take a weighted average of the excess death rates in each of the cells, weighted by the 2019 death count.

\subsection{Statistical model}

To assess overall excess death rates during the COVID-19 pandemic, and to adjust differences in excess death rates by political party affiliation for age bins and county of residence, we used multivariable linear regression models. In some cases, we ran these regressions separately by time period. We selected April 5, 2021 as a cutoff for when vaccines were widely available to adults in Florida and Ohio. The federal deadline for when states had to open eligibility for vaccines for all adults was April 24, 2021, but Florida and Ohio did so earlier. Florida made vaccines available to all adults on April 5, 2021. For more details on Florida’s policy see: \url{https://floridahealthcovid19.gov/latest-vaccine-updates/}. Ohio made vaccines available to all adults earlier, on March 29, 2021. For more details on Ohio’s policy see: \url{https://governor.ohio.gov/media/news-and-media/expanded-vaccine-eligibility-cleveland-mass-vaccination-clinic-opens-to-public-03162021}. Hence, we selected the latter of the two dates, April 5, 2021, as the date by which all adults in both states were eligible for vaccines. Because our data is monthly, we use April 2021 data on and onwards to reflect this period, which corresponds closely to when vaccines were widely available.

To estimate the difference in excess death rates by month (the third panel of Figure 1), regression models were of the form:

\begin{equation*}
  Y_{tcpa} = \gamma_{tca} \times W_{tca} + \beta_{t} \times 1(\text{Affiliation = Republican})_{p} \times 1(\text{Year-month} = t)_{t} + \epsilon_{tcpa}
\end{equation*}

where $W_{tca}$ is a set of fixed effects for year-month-by-county-by-age-bin, and we plot the $\beta_{t}$ coefficients over time (marking the relative difference in excess deaths for Republicans).

We also estimated regression models that pooled data from multiple months and recovered a single coefficient for three different periods of the COVID-19 pandemic: April 1, 2020 – December 31, 2021; April 1, 2020 -- March 31, 2021; and April 1, 2021 -- December 31, 2021. When estimating overall excess death rates, these regressions were simply estimates of the constant in a regression:

\begin{equation*}
  Y_{tcpa} = \beta_{0} + \epsilon_{tcpa}
\end{equation*}

where the coefficient of interest is $\beta_{0}$.

To estimate the effect of vaccination rates on the difference in excess death rates by month (the third panel of Figure 1), regression models were of the form:

\begin{align*}
  Y_{tcpa} &= \gamma_{tca} \times W_{tca} + \beta_{t} \times 1(\text{Affiliation = Republican})_{p} \times 1(\text{Year-month} = t)_{t}\\
  &+ \tau_{t} \times 1(\text{Affiliation = Republican})_{p} \times 1(\text{Year-month} = t)_{t} \times \text{CountyVaccionationRate}_{c} +  \epsilon_{tcpa}
\end{align*}

where $W_{tca}$ is a set of fixed effects for year-month-by-county-by-age-bin, $CountyVaccineRate_{c}$ is the share of the county’s population that had at least one dose of vaccines as of June 2021, and we plot the $\tau_{t}$ coefficients over time (marking the change in the relative difference in excess deaths for Republicans when increasing county vaccination by one standard deviation (~10 percent)).

In all regression cases, we cluster standard errors by county.

In the last Figure, we calculate the average excess death rates by county for Democrats and Republicans between April 2020 to March 2021 and April 2021 to December 2021.  We plot the county level measures and fit a quadratic curve for each group and time period. We also plot binned means using the binsreg package provided by Cattaneo et al. (2021). 

\end{appendix}
\end{document}